\documentclass[journal]{IEEEtran}
\usepackage{amsmath,amsfonts}
\usepackage{algorithmic}
\usepackage{algorithm}
\usepackage{array}
\usepackage[caption=false,font=normalsize,labelfont=sf,textfont=sf]{subfig}
\usepackage{textcomp}
\usepackage{stfloats}
\usepackage{url}
\usepackage{verbatim}
\usepackage{graphicx}
\usepackage{cite}
\usepackage{tabularx}
\usepackage{xspace}
\usepackage{booktabs}
\usepackage{color}
\usepackage{hyperref}

\usepackage[svgnames]{xcolor}
\usepackage{tcolorbox}
\usepackage{marginnote}
\usepackage{adjustbox}
\newif\ifrevise
\revisetrue
\global\marginparsep=10pt
\global\marginparwidth=24pt

\begin{document}

\newcommand{\eg}{e.g.}
\newcommand{\ie}{i.e.}
\newcommand{\etal}{et al.}
\newcommand{\etc}{etc.}
\newcommand{\systemname}{\textit{SynthLens}\xspace}
\newcommand{\Ex}[1]{\rm E_{#1}\xspace}
\newcommand{\Px}[1]{\rm P_{#1}\xspace}
\newcommand{\blue}[1]{\textcolor{Black}{#1}}


\title{\systemname: Visual Analytics for Facilitating Multi-step Synthetic Route Design}

\author{Qipeng Wang, Rui Sheng, Shaolun Ruan, Xiaofu Jin, Chuhan Shi, Min Zhu
\thanks{Q. Wang, M. Zhu are with Sichuan University. Email: wangqipengscu@stu.scu.edu.cn, zhumin@scu.edu.cn 
\par R. Sheng, X. jin are with Hong Kong University of Science and Technology. Email: {rshengac, xjinao}@connect.ust.hk
\par S. Ruan is with Singapore Management University. Email: slruan.2021@phdcs.smu.edu.sg
\par C. Shi is with Southeast University. Email: chuhanshi@seu.edu.cn
\par Corresponding author: M. Zhu and C. Shi.}
\thanks{Manuscript received April 19, 2021; revised August 16, 2021.}}

\markboth{Journal of \LaTeX\ Class Files,~Vol.~14, No.~8, August~2021}%
{Shell \MakeLowercase{\textit{et al.}}: A Sample Article Using IEEEtran.cls for IEEE Journals}


\maketitle

\begin{abstract}
Designing synthetic routes for novel molecules is pivotal in various fields like medicine and chemistry. 
In this process, researchers need to explore a set of synthetic reactions to transform starting molecules into intermediates step by step until the target novel molecule is obtained. 
However, designing synthetic routes presents challenges for researchers. 
First, researchers need to make decisions among numerous possible synthetic reactions at each step, considering various criteria (\eg, yield, experimental duration, and the count of experimental steps) to construct the synthetic route. 
Second, they must consider the potential impact of one choice at each step on the overall synthetic route.
To address these challenges, we proposed \systemname, a visual analytics system to facilitate the iterative construction of synthetic routes by exploring multiple possibilities for synthetic reactions at each step of construction. 
Specifically, we have introduced a tree-form visualization in \systemname to compare and evaluate all the explored routes at various exploration steps, considering both the exploration step and multiple criteria.
Our system empowers researchers to consider their construction process comprehensively, guiding them toward promising exploration directions to complete the synthetic route. We validated the usability and effectiveness of \systemname through a quantitative evaluation and expert interviews, highlighting its role in facilitating the design process of synthetic routes. Finally, we discussed the insights of \systemname to inspire other multi-criteria decision-making scenarios with visual analytics. 
\end{abstract}


\section{Introduction}
The synthesis of novel molecules is crucial to many fields, ranging from drug design to material science. 
For example, in drug design, a crucial step in successfully synthesizing safe and effective drugs is designing synthetic routes that can produce complex molecules and directly impact the efficiency and viability of the synthetic process.
These routes enable chemists not only to discover but also to produce novel molecules of drugs. 
However, it is a challenging task to design a feasible synthetic route, often taking over a decade and costing upwards of US\$100 million. 

Currently, researchers typically use retrosynthesis \cite{retro1988corey}, an automated method that breaks down the target molecule into starting molecules, to construct synthetic routes. However, they often find these routes impractical and prefer to rely on reference papers. Therefore, they explore synthetic reactions from reference papers to manually design synthetic routes transforming starting molecules into intermediates step by step until obtaining the target molecule.
At each step of the designing process, researchers must balance multiple criteria (\eg, yield, duration, and experimental difficulty), which may conflict when selecting the optimal synthetic reaction.
Such multi-criteria decision-making task makes the synthetic route design process tedious and time-consuming. 
Additionally, each decision can affect not only current step's outcome but also subsequent choices and the overall synthetic route. 
For instance, choosing the former between two possible reactions at the early step: a high-yield but slower versus a low-yield but fast reaction may improve yield but extend the overall duration, leading to increased costs and delays. The decision may also restrict reagent options in later steps due to intermediate stability. 
Additionally, the reaction at this early step may limit the choice of reagents in subsequent steps due to the stability of the intermediates produced. 
The complexity is further heightened by the inclusion of qualitative factors (\eg, experimental procedures) in addition to quantifiable numerical values.

Several visualization tools can help users make multi-criteria decisions. 
For example, FSLens\cite{Chen2024FSLens} can assist users in making decisions regarding the location of new fire stations by considering factors such as distance and time. 
WarehouseVis\cite{Li2020Warehouse} aims to optimize and simplify the decision-making process for warehouse location selection according to heterogeneous data like renting costs, reachability, and traffic conditions. 
In the context of designing synthetic routes, where a decision sequence from the starting molecule to the target molecule involves a series of synthetic reactions, each with multiple experimental procedures, the complexity of the decision sequences requires a more comprehensive approach to ensure a successful design. 
However, current tools cannot aid users in decision-making among numerous possible choices at each step while considering the influences on subsequent choices. 
Additionally, they fail to provide users with a comprehensive view of decision sequences. 
To address the mentioned challenges, we propose \systemname, a visual analytics system that assists researchers in designing synthetic routes by integrating and comparing potential synthetic reactions effectively. 
Firstly, \systemname facilitates designing a synthetic route from the starting molecule to the target molecule by integrating synthetic reactions, whose information is extracted from papers. 
Secondly, \systemname helps users make a series of decisions that is sequential decision-making by providing an overview of the decision sequences in a comprehensive visualization. 
Thirdly, \systemname provides users with multiple choices at each step of constructing sequence decisions and allows them to compare all decision sequences considering multiple factors. 
Consequently, \systemname enables users to make informed decisions and identify the most promising decision sequences for exploration.

In conclusion, the contributions of our work are:

\begin{itemize}
    \item The system design requirements are summarized by cooperating with six chemistry experts who are dedicated to designing the synthetic routes of novel molecules.
    \item A visual analytics system that can help chemists make sequential decisions considering subsequent impacts in designing synthetic routes. To be specific, we introduced a comprehensive visualization to compare and evaluate each decision or all constructed sequences at various steps of exploration, considering multiple criteria.
    \item Two case studies and expert interviews to validate the effectiveness of our system. Specifically, we found the workflow of \systemname can be extended to multiple field tasks such as retrosynthesis. 
\end{itemize}

\section{Problem Formulation}
\begin{figure}
    \centering
    \includegraphics[width=\linewidth]{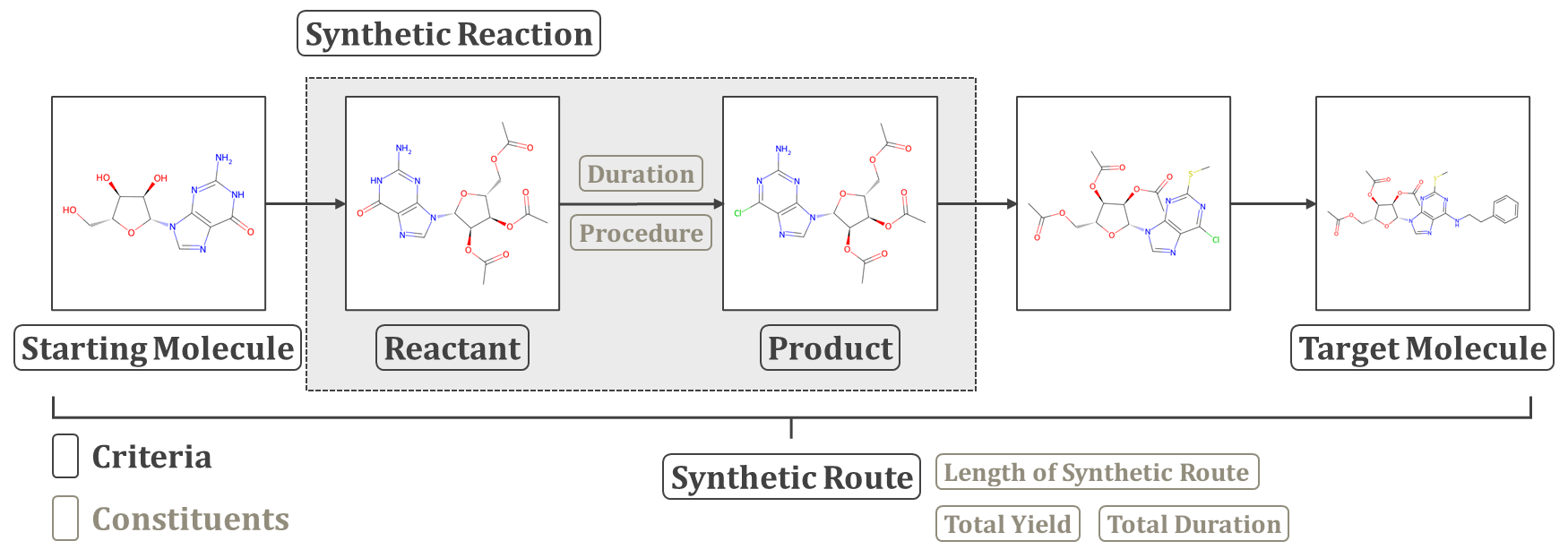}
    \caption{A synthetic route from the starting molecule to the target molecule consists of several synthetic reactions, each involving specific reactants and products. The evaluation of a synthetic reaction contains criteria such as duration and experimental procedure.}
    \label{fig:figure2}
\end{figure}

To clarify the complex design challenge of constructing synthetic routes, we have clearly defined several key terms that are extracted from our comprehensive literature review in the following content, which are illustrated in Fig.\ref{fig:figure2}.

\begin{itemize}
    \item A \textbf{synthetic route} is a sequence of synthetic reactions that can transform a starting molecule into a target molecule\cite{retrosynthesis2022ishida}. We refer to the synthetic route that has not yet successfully achieved the target as an \textbf{intermediate synthetic route}.
    \item \textbf{Starting molecules} are the starting compounds in a synthetic route\cite{Finnigan2021definiation}. Experts usually determine the starting molecule based on their previous research experience. \textbf{Target molecules} represent novel molecules that experts endeavour to synthesize, given that these molecules are hitherto unavailable and have not been synthesized previously\cite{Finnigan2021definiation}.
    \item A \textbf{synthetic reaction} is a chemical process involving bond cleavage and formation, by which two or more molecules react to form a more complex molecule\cite{Finnigan2021definiation}. 
    \item The \textbf{reactant} of a synthetic reaction refers to the molecule that acts as the starting material\cite{gerasimchuk2022chemical}, which undergoes a chemical change during the reaction, breaking existing chemical bonds and forming new ones to generate the \textbf{products}\cite{gerasimchuk2022chemical}.
    \item The \textbf{main chain} of a molecule is the basic skeleton and serves as the basis of organic molecules\cite{Huang2016mainchain}. \textbf{Side chains} of a molecule refers to the components attached to the main chain\cite{Li2022sidechain}, mainly determining the properties of a molecule.
    \item \textbf{Retrosynthesis} involves breaking down the structure of a target molecule into simpler, more easily synthesizable reactant, thereby completing the synthesis route\cite{retro1988corey}.
\end{itemize}

Designing a \textbf{synthetic route} is the process of identifying various \textbf{synthetic reactions} and integrating them sequentially, with the goal of transforming the \textbf{starting molecule} into \textbf{intermediate products} step by step until obtaining the desired \textbf{target molecule}. 
To construct a synthetic route, experts need to identify and assess various synthetic reactions from different papers. 
Then, for each step of the design process, they may filter multiple candidate reactions that have the potential to make up the whole synthetic route. 
To assess these candidate reactions, experts consider multiple criteria, including \textbf{reaction conditions}, \textbf{yield}, \textbf{reaction duration}, and \textbf{synthetic step count}. We will detail these criteria as follows.

\begin{itemize}
    \item \textbf{Reaction conditions} are the various parameters that control the progress of a chemical reaction, such as temperature, pressure, solvent, and catalyst\cite{kwon2022condition}. The difficulty of the reaction conditions is directly reflected in the complexity of carrying out the \textbf{experimental procedures}.
    \item \textbf{Yield} is a measure of the efficiency of one chemical reaction, representing the ratio of the actual amount of product obtained in an experiment to the theoretically calculated amount\cite{Schwaller2020yield}. Improving yield is a significant objective as it directly impacts both the efficiency and cost-effectiveness of the reaction.
    \item \textbf{Reaction duration} refers to the time a synthetic reaction takes to complete\cite{Xu2019duration}. Experts can optimize the reaction duration to speed up the overall synthetic route, enabling the fast production of the products and reducing the required material.
    \item \textbf{Synthetic step counts} is the number of reaction steps required to transform the starting molecule to the target molecule. Shortening the synthetic route may decrease the potential side reactions and impurities, leading to a more effective synthesis.
\end{itemize}

\section{Related Work}

\subsection{Multi-step Synthetic Route Design}
Computer-Aided Synthesis Planning (CASP) aims to assist researchers in designing and optimizing multi-step synthetic routes by using computer technology. 
The goal of CASP is to automate the design of synthetic routes, reduce the trial-and-error process, and increase the efficiency and success rate of synthesis. 
Firstly, several studies on CASP can generate a variety of synthetic routes using predictive models. 
For example, WODCA\cite{Gasteiger2000WODCA} is a tool for forward synthetic and retrosynthetic route prediction using the fundamental nature of chemical bonding to guide suitable retrosynthetic breakpoints. 
AiZynthFinder\cite{Genheden2020AiZynthFinder} is open-source software that uses an artificial neural network policy to quickly and efficiently perform retrosynthetic prediction, providing potential precursors for a given molecule. 
Nevertheless, the accuracy of predicted routes can sometimes fall short of expectations. 
Consequently, human expertise should be integrated to improve the reliability and precision of prediction. 
Some tools provide intuitive functions for the hands-on construction and iterative optimization of synthetic routes. 
ICSYNTH\cite{hans2013ICSYNTH} utilizes reaction cores extracted from various databases to construct synthetic suggestion trees under user control. 
LinChemIn\cite{Pasquini2023LinChemIn} is a Python toolkit that enables chemo-informatics operations on synthetic routes and reaction networks, allowing users to operate synthetic routes. 
However, existing studies lack the interaction for users to add synthetic reactions into their synthetic routes dynamically, which aligns with the decision process of chemists. 
Furthermore, current tools face challenges in facilitating researchers in choosing the most suitable synthetic routes from multiple candidates by considering various criteria, such as yield, duration, experimental procedures, and reaction conditions.

\subsection{Visual Analytics for the Chemical Domain}
Visual analysis has been widely employed in various fields in the chemical domain, such as drug discovery\cite{medina2008chemicalspace}, chemistry education\cite{wu2001class}, and chemical reaction exploration\cite{Chiu2014reaction}. 
Some of these studies can facilitate the discovery of novel molecules and identify relationships between different molecules. 
For instance, DataWarrior\cite{sander2015datawarrior} can assist users in visualizing and analyzing chemical biology data and identifying correlations and associations among various molecules. 
Naveja\etal\cite{naveja2019finding} presents the novel constellation plots to enable users to identify and interpret Structure-Activity Relationships (StARs). ChemoGraph\cite{kale2023chemograph} allows users to explore novel molecules in the form of hypergraphs. 
Furthermore, other studies can support users in leveraging visual analysis technology to study synthetic reactions or routes by enhancing the learning and understanding of complex chemical processes. 
For example, Chiu \etal\cite{Chiu2014reaction} utilized enhanced visual units to help students develop a deeper understanding of chemical reactions by connecting and refining knowledge. 
Moreover, RetroLens\cite{shi2023retrolens} can provide the retrosynthetic routes of a target molecule and present candidate correction steps that AI recommends to promote collaboration between humans and AI. 
However, our work differs from all the prior studies in that the goal of our work is to assist researchers in designing synthetic routes based on numerous potential synthetic reactions, which is a complex multi-step decision-making problem and requires researchers to balance various criteria in the process.

\subsection{Visual Analytics for Multi-criteria Decision-making}
Common single-criteria decision-making tasks focus on only one objective, such as optimizing costs or maximizing benefits\cite{brumar2024task}, while multi-criteria decision-making refers to identifying a satisfactory choice from a range of options evaluated against several criteria. 

Furthermore, numerous multi-criteria decision-making systems have been proposed to cater to the unique requirements of specific domains such as location selection \cite{liu2017smartadp, Li2020Warehouse, Chen2024FSLens} and route planning\cite{Partl2016Pathfinder, weng2021bnva, rauscher2024ski}.
Route planning is a typical multi-criteria decision-making problem that involves evaluating and prioritizing options based on multiple criteria\cite{shin2023route}. 
Pathfinder\cite{Partl2016Pathfinder} is designed to facilitate users with interactive capabilities for querying, ranking and comparing paths within large and multivariate network datasets.
Weng \etal\cite{weng2021bnva} designed BNVA that proposed a progressive route decision-making strategy to evaluate the performance of bus routes.
SkiVis \cite{rauscher2024ski} enables users to receive customized route recommendations based on preferences such as steepness and crowdedness.
In terms of location selection issues, WarehouseVis\cite{Li2020Warehouse} combines visual analytics and interactive modelling to optimize and simplify the decision-making process for warehouse siting in retail logistics management. 
SmartAdP\cite{liu2017smartadp} facilitates the comparison of multiple billboard placement solutions
by multiple attributes, such as traffic volume, speed, origins and destinations.
FSLens\cite{Chen2024FSLens} integrates fire records and collaborative decision-making technology to identify and improve the siting of fire stations. 
There are also various studies targeting specific multi-criteria decision-making approaches rather than particular applications or domains. WeightLifter\cite{pajer2017weightlifter} enhances multi-criteria decision-making by enabling the exploration of weight spaces for up to 10 criteria, improving the decision efficiency and credibility. Huang \etal\cite{huang2024evo} proposed a visual analytics framework to explore and compare the evolutionary processes in evolutionary multi-criteria optimization algorithms.

However, in designing chemical synthetic routes, researchers must consider a variety of quantitative factors (\eg, yield and reaction duration) and qualitative textual data in a long, multi-stage decision sequence, which poses significant challenges.
Therefore, it is crucial to develop a new visual analytics system to streamline the extensive decision-making process involved in designing these routes.

\section{Design Study}\label{sec:sec3}
We collaborated with six experts ($\Ex{1}$-$\Ex{6}$) from the field of organic chemistry, each with varying years of experience and specialized in different research areas. 
We gained insights into their workflows and challenges in designing routes for novel molecules through semi-structured interviews, each lasting around 60 minutes.
Additionally, we held biweekly meetings with experts to ensure our system aligned with their domain needs, iteratively refine the system based on their immediate feedback. 
This collaboration helped us identify the crucial factors influencing their decision-making process and the five key design requirements for our system. 

\begin{itemize}
    \item[\textbf{R1}] \textbf{Specify the starting molecule and required synthesis reaction.}
    The experts need to specify the starting molecule of the synthesis and search for potentially expected synthetic reactions from papers based on their expertise and knowledge. 
    Therefore, our system should provide a flexible input interface for this purpose.

    \item[\textbf{R2}] \textbf{Examine the details of synthetic reaction.}
    Experts need to review numerous papers to find synthetic reactions that fulfil their requirements at the step, such as experimental procedures and specific products. For example, $\Ex{5}$ has emphasized, \textit{``Papers often include extensive details such as experimental procedures, flowcharts, molecule structure diagrams, and spectra diagrams, making it time-consuming to identify useful reactions.''} Therefore, it is crucial to help experts effectively assess these details in each paper and decide whether to integrate a particular synthetic reaction into their synthetic route. 

    \item[\textbf{R3}] \textbf{Construct synthetic routes based on synthetic reactions.}
    Experts will construct synthetic routes step by step through several identified synthetic reactions. Specifically, they might explore multiple potential synthetic reactions at each step. Therefore, our system should support users in manually constructing synthetic routes.

    \item[\textbf{R4}]  \textbf{Support synthetic route comparison at various steps.} 
    $\Ex{6}$ has highlighted that experts frequently assess the intermediate synthetic routes to choose the optimal one for further exploration in the middle step of the designing process. Furthermore, experts also need to compare all completed routes to make the final decision. To facilitate this crucial comparison process, our system should provide a flexible approach to support experts in comprehending multiple factors such as the synthetic step count, yield, duration, and experimental step details at any step. Besides, several experts have mentioned that experts may prioritize differently for various synthetic tasks. Therefore, our system should allow users to adjust the weight of different criteria based on their specific needs and preferences for ranking candidates. 

    \item[\textbf{R5}] \textbf{Optimize synthetic routes iteratively.}
    Experts often need to optimize the constructed synthetic routes to achieve their desired criteria. Our system should support iterative optimization by allowing experts to modify and refine the synthetic routes based on their assessment and analysis iteratively.
\end{itemize}

\section{Data Processing}\label{sec:sec4}
In this section, we first present the overview of our system, followed by a detailed introduction to the paper access modules and the process of information extraction from papers. Then, we propose the workflow of constructing the synthetic route.

\subsection{System Overview}
\begin{figure}
    \centering
    \includegraphics[width=\linewidth]{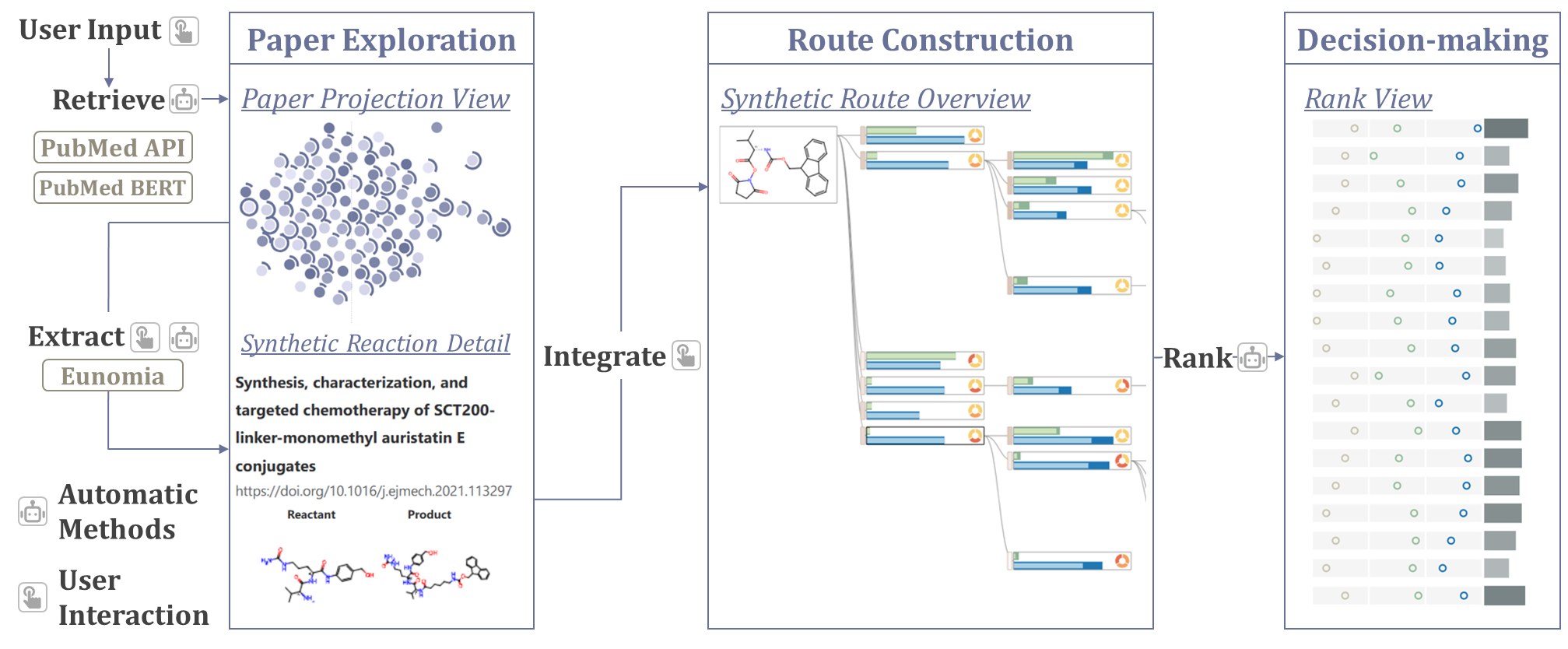}
    \caption{The analyzing workflow of \systemname: \textbf{User Input}: a user can define a starting molecule and expected synthetic reaction, then manually select one from \textit{retrieved papers} for exploration. \textbf{Information Extraction}: the synthetic reaction details are then \textit{extracted automatically}. \textbf{Synthetic Route Construction}: the user can choose to \textit{integrate} the synthetic reaction into the synthetic route construction to form various decision sequences. \textbf{Decision-making}: finally, the user can select the optimal synthetic route from computed \textit{rankings} supplemented by his own preferences. The whole process is the combination of \textit{automatic methods} and \textit{user interaction}.}
    \label{fig:figure3}
\end{figure}

We present \systemname, a visual analytics system that assists organic chemists in designing synthetic routes by integrating synthetic reactions extracted from papers, enhancing the route design process. 
The data process of implementing \systemname contains two modules: paper access and analysis, and information extraction (Fig.\ref{fig:figure3}).
\systemname can retrieve papers using the molecule that the users input as search terms. 
Then, \systemname analyzes and processes the retrieved papers to demonstrate the distribution of these papers. 
In the information extraction module, we use an LLM-based information extraction technique to extract information on synthetic reactions from papers and transform them into structured data.
As for front-end visualization, \systemname allows the users to compose these reactions into synthetic routes and visualize these routes in a tree-form visualization. 
\systemname also provides views to assist the users in exploring the decision sequences and making decisions in the designing process.

\subsection{Papers Access and Analysis}
Users usually search for papers related to the reactants of each synthetic reaction. The amount of papers retrieved may be substantial, and each paper may propose multiple synthetic reactions. Therefore, it can be challenging for users to identify which paper is relevant to their requirements and what sections in a paper contain useful information. To address this, \systemname needs to retrieve the papers using specific molecules as search terms and extract details about specific synthetic reaction from the chosen paper.

Requesting the user to enter the name of the starting molecule is not advisable because the molecule's name can be complex and may not accurately correspond to the correct molecular structure.
As a result, \systemname allows users to set the starting molecule by drawing the structure of a molecule. 
We have incorporated the open-source chemical structure editor Ketcher\footnote{https://github.com/epam/ketcher} into our system to assist users in drawing molecular structures. This editor can generate SMILES (Simplified Molecular Input Line Entry System) strings of a molecule once the user has drawn the molecular structure. SMILES can convey the accurate molecule structure in a compact and machine-readable format. 
We search papers using API provided by PubMed\cite{pumbedapi}, which accepts SMILES strings as search terms. We can retrieve a list of relevant papers that include information such as title, abstract, keywords, digital object identifiers (DOI), and citation counts by this API. 

As the users expect our system to categorize the retrieved papers according to the distribution of their research content, we transform the abstracts of the papers into vector embeddings to capture the semantic information of the text and map semantically similar abstracts to proximity locations in embedded space. To achieve this, we use the PubMedBERT model\cite{gu2021pubmedbert}, which is a pre-trained language model designed for the biomedical domain, to convert the paper title and abstracts into high-dimensional embeddings. 
Then we use the t-SNE algorithm \cite{maaten2008tsne} to project high-dimensional embeddings into two-dimension points in the 2D plane, providing a spatial representation of papers. The positioning of points provides a visual cue for the semantic similarities among papers, helping users intuitively discover potential relationships that may not be apparent in tables.

\subsection{Information Extraction from Papers}\label{chapter_4_2}
After users choose a paper to read, they may encounter difficulty in extracting a specific synthetic reaction. This is because one paper may propose multiple synthetic reactions and each synthetic reaction may contain various details. Therefore, our system aims to automate the extraction process. To achieve this, we first use the API provided by UnpayWall\footnote{https://unpaywall.org/products/api} to obtain a PDF file of the paper.

To extract synthetic reactions details from the paper, we use Eunomia\cite{ansari2023eunomia}, a chemist AI agent. This tool adopts the commonly used workflow for information extraction using LLMs\cite{paperqa2023lala}. Firstly, it converts both papers and queries into text embedding, which are then stored within an embedding database. It then conducts semantic similarity searches to find the most relevant paragraphs.
Furthermore, Eunomia employs Chain-of-Verification\cite{cov2024dhuliawala} to reduce hallucinations in LLMs. Base on this automatic extraction pipeline, we created the prompt template in which \textit{\textbf{A}} refers to the reactant of the reaction, \ie, the search term and \textit{\textbf{B}} refers to the specific expected synthetic reaction:

\begin{itemize}
    \item[] \textit{1) You are an expert chemist. This document describes the synthetic route or synthetic reaction of the \textit{\textbf{A}}.}
    \item[] \textit{2) Find the information of the specific reaction \textit{\textbf{B}} and the reactant of the reaction must be \textit{\textbf{A}}.}
\end{itemize}

Then, we asked Eunomia to provide the answer, including the following items: reactants, products, solvent, reagent, catalysts, duration, instruments, operation, and yield, and we specified the output format:

\begin{itemize}
    \item[] \textit{Your final answer should be a structured JSON format including these items: reactants, products, solvent, reagent, catalysts, duration, instruments, operation, and yield. The answer should be "null" if you cannot find the expected reaction.} 
\end{itemize}

\begin{figure}
    \centering
    \includegraphics[width=\linewidth]{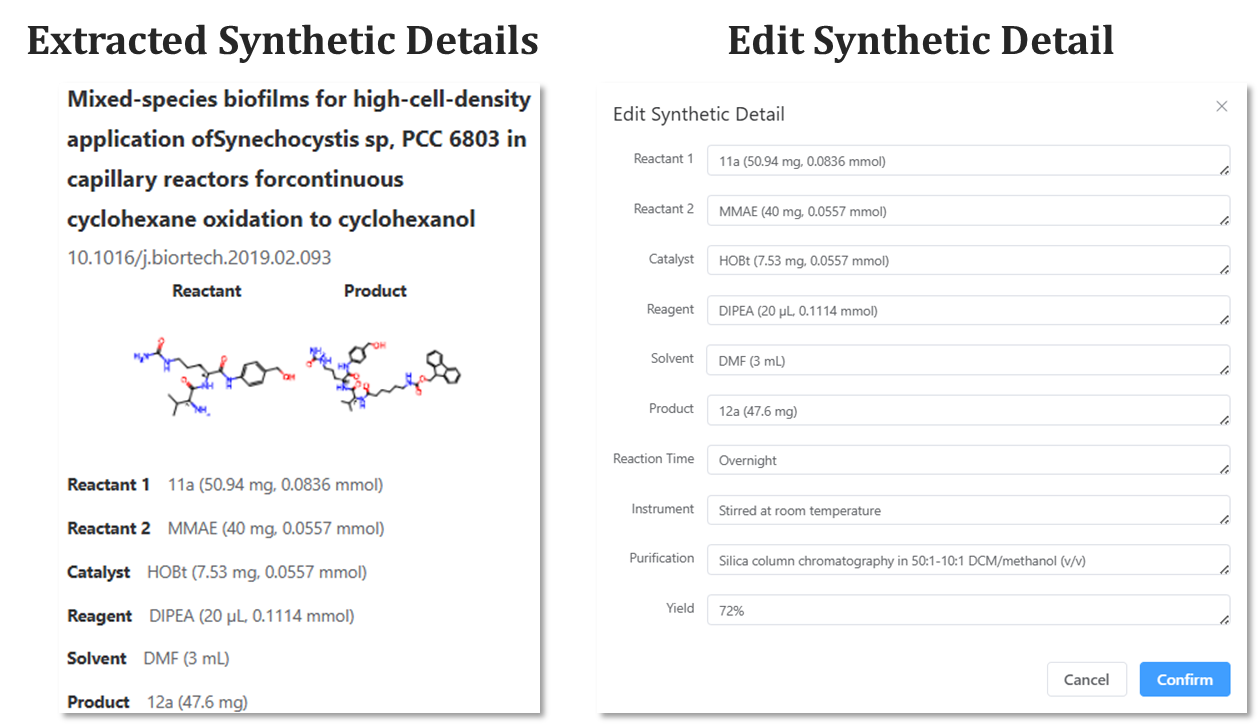}
    \caption{
    The extracted synthetic details are presented in a table format in our system, consisting of several parts: \textbf{\textit{raw materials}}, \textbf{\textit{experimental operations}}, \textbf{\textit{duration}}, and \textbf{\textit{yield}}, etc. Moreover, our system allow users to modify the extracted details within a form.}
    \label{tab:table1}
\end{figure}

We specified all the expected reactions as parameter \textit{\textbf{B}} and performed several extractions. One example of the extracted information of a synthetic reaction is presented in Fig.\ref{tab:table1}. 

Although Eunomia claims that it is able to sace the agent's output in JSON format, we used the JSON module in Python to parse the model's output string in order to ensure that the output format fully meets our needs. If parsing fails, we retry by inserting a directive that asks the model to strictly adhere to the prompt following the original prompt until the maximum time of retrying is reached.

Additionally, we provide users with a key metric: \textit{Context Relevancy}\cite{ragas2023Shahul} to help them evaluate the extracted information. \textit{Context Relevancy} is an indicator used to evaluate the performance of LLMs in retrieving information, and it measures the relevancy between the content of the answer and its context.
Also, to avoid over-reliance on our system, we added disclaimers in SynthLens to remind users that there may be inaccuracy in the extracted information because of technological limitations. Meanwhile, we provide a link to each paper, allowing them to access papers online directly to verify extracted synthetic details.

\subsection{Synthetic Route Construction}

While there are automatic methods for generating synthetic routes like retrosynthesis\cite{retro1988corey}, experts caution that the routes automatic generated may not be feasible in practice due to complexity and experimental challenges of these routes. Instead, they prefer designing synthetic routes based on information from reference papers. Therefore, \systemname is designed to provide convenient access to valuable papers and directional suggestions, which helps reduce errors that could arise from full automation and construct synthetic routes.
To facilitate the construction process, we defined a tree-form data structure to organize decision sequences consisting of synthetic reactions from the starting molecule to the target molecule, comprising the following elements.

\begin{itemize}
    \item The \textbf{root} of a tree represents the starting molecule of the synthetic route that users specified. After setting the starting molecule, \systemname will automatically retrieve a number of papers using the starting molecule as a search term.
    \item A \textbf{node} in the tree represents a synthetic reaction extracted from retrieved papers, whose reactant is the product of the \textbf{parent node}. 
    Each node also contains essential information, including yield, duration, and SMILES strings of the reactants and products.
    Upon choosing a node, \systemname will automatically retrieve papers using the product of this node as the search term. 
    The synthetic reaction extracted from specific retrieved papers can be integrated as subsequent nodes, \ie, \textbf{child nodes}. 
    We also recorded the nodes' total yield and duration because these values will accumulate as the reaction proceeds sequentially. The output of one step serves as the input for the next. To be specific, if the yield of the first step is 80\% and the yield of the second step is 90\%, then the second step yields 72\% totally (\ie, \(80\% \times 90\% = 72\%\)). 
    Similarly, we obtain the total duration by calculating the sum of the duration of this node and the parent node.
    \item A \textbf{leaf node}, typically representing the product molecule, has no child nodes. Nodes that are neither root nor leaf are called \textbf{intermediate nodes}, which represent intermediate molecules. The node's \textbf{layer count} refers to the number of synthetic reactions from the root node(\ie, the starting molecule) to the current node(\ie, the intermediate molecule).
    \item \textbf{Deep and bread exploration strategies.} Deep exploration strategy involves going as far down a branch of the tree as possible before backtracking to explore new branches, while bread exploration strategy involves exploring all neighbour nodes at the present step before moving on to nodes at the next level.
\end{itemize}

\section{Visual Design}

\begin{figure*}
      \centering
  \includegraphics[width=\linewidth]{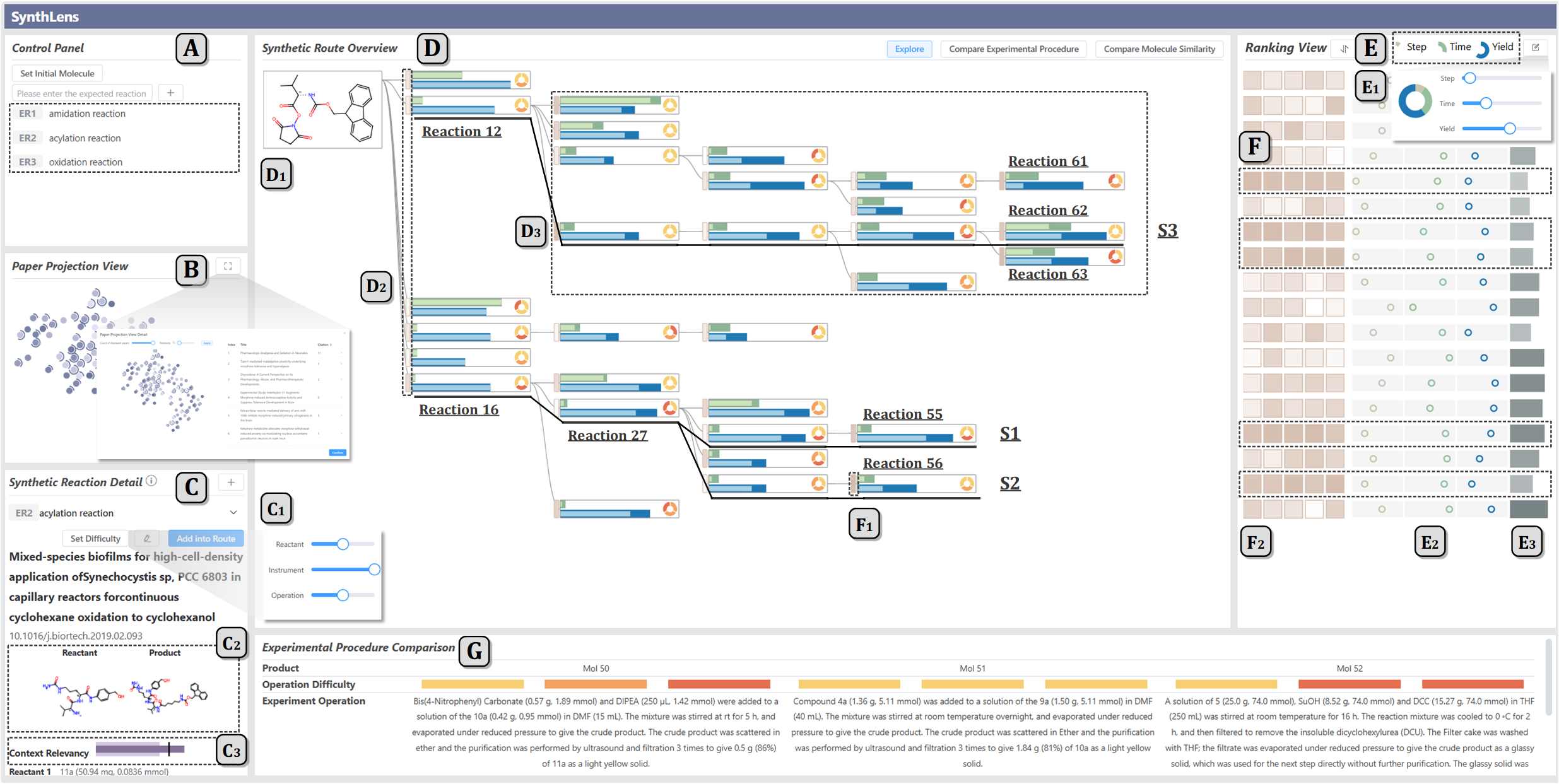}
  \caption{
    \systemname: (A) The Control Panel allows users to specify the starting molecule and potentially expected synthetic reactions before designing the synthetic routes. 
    (B) The Paper Projection View presents the distribution of retrieved papers. 
    (C) The Synthetic Reaction Detail shows the detail of extracted synthetic reactions. 
    (D) The Synthetic Route Overview presents a tree-form visualization of the decision sequences of the synthetic routes. 
    (E) The Similarity View shows the similarity in structure among specific molecules. 
    (F) The Rank View visualizes the rank of decision sequences considering three factors with flexible weights. 
    (G) The Experimental Procedure Comparison assists users in comparing the experimental procedures of multiple synthetic reactions.
  }
  \label{fig:figure1}
\end{figure*}

In this section, we detail the visual design of \systemname, which leverages a combination of manual inputs and automated processes to meet the design requirements specified in Section \ref{sec:sec3}. 
As Fig.\ref{fig:figure1} shows, users can set a starting molecule and input the expected synthetic reaction in the Control Panel (\textbf{R1}).
Then, they can manually select a paper in the Paper Projection View, where the system has already performed searches and computations.
The experimental procedures for the synthetic reactions are then automatically extracted and presented in the Paper Detailed View (\textbf{R2}). 
After annotating the difficulty of a synthetic reaction's experimental procedure, the user can integrate the reaction into the Synthetic Route Overview and end up producing multiple decision sequences (\textbf{R3}, \textbf{R5}). 
After route construction, the user can manually choose the most appropriate decision sequence from the automatically computed rankings in the Rank View (\textbf{R4}).

\subsection{Paper Overview}
To help users explore papers related to a synthetic reaction and decide whether to integrate the reaction into the decision sequence, we proposed the Paper Overview. 
This view consists of two sub-views: the Paper Projection View and the Paper Detail View. 
Specifically, the Paper Projection View provides a comprehensive overview of retrieved papers with diverse distribution of research content, while the Synthetic Reaction Detail presents the details of a particular synthetic reaction extracted from the corresponding paper.

\subsubsection{Control Panel}
The Control Panel (Fig.\ref{fig:figure1}(A)) allows users to draw the structure of the starting molecule of the whole synthetic route in the editor. 
Users can plan the synthetic route by identifying the expected reactions to transform the starting molecule into the target molecule based on their differences and chemical domain knowledge. 
So, the Control Panel allows users to input these expected reactions. 
This assists the following process of extracting synthetic reactions.

\subsubsection{Paper Projection View}

The retrieved papers present a diverse distribution in research content and include detailed information like titles, abstracts, citation counts, keywords, \etc, making the exploration of a large number of papers a challenging task. After interviewing experts, we learned they usually rely on heuristic exploration methods to identify target papers for further exploration. This approach allows users to precisely determine the most relevant papers based on a deep understanding of their contents. Therefore, we employ a scatter plot in the Paper Projection View (Fig.\ref{fig:figure1}(B)), which helps users find semantically similar papers after identifying one paper of interest based on their expertise. For example, users are more likely to explore papers spatially close to the selected paper in the scatter plot, rather than just focusing on those ranking highest in the paper list.

\textbf{Description.}
In the scatter plot of the Paper Projection View, each point represents a retrieved paper, with the color intensity indicating the paper's relevance to the search term. Rather than using point size to represent relevance, color intensity ensures points remain visible even when the paper’s relevance to the search term is low. In contrast, smaller points might become indistinguishable at a smaller scale.
After interviewing experts, we found that they prefer manually exploring papers. This preference arises because automated filters can not capture the semantic difference between papers and carry the risk of omitting significant papers. 

After transforming the abstract of these papers into embeddings and dimensionality reduction introduced in Section \ref{sec:sec4}, we obtained the coordinates of each paper projected in the 2D plane. 
However, if these points are mapped directly by coordinates, there will be an overlap among these points.
To reduce the overlapping of the points in the scatter plot, we employed a force-directed algorithm provided by d3.js\cite{Bostock2011d3} that applies a repulsive force to maintain at least a minimum distance between points, without changing their relative positions.
In addition, we added an arc surrounding each point encoding the citation count of this paper, which can indicate its credibility.
Proximity in distance among points indicates a similarity in research content. 
Users are willing to prioritize papers that not only demonstrate a high degree of relevance to the key molecule but also have high citation numbers. 
Upon exploring a paper, subsequent exploring choices should focus on papers that are near to ensure the consistency of exploration. We display the 100 papers retrieved as related to the specified molecule in the Paper Projection View because experts usually retrieve less than 100 relevant papers when searching synthetic reactions. Showing too many can overwhelm the analysis with unrelated papers.
If the user finds the points too dense, they can zoom in and drag to see a clearer picture of the scatter's positional relationships.

\textbf{Interaction.}
Hovering over a point triggers a tooltip to appear, presenting the details of the corresponding paper, including title, abstract, DOI, citation number, and keywords. 
These details can guide the users' decision on whether to explore a specific paper.
When a point is clicked, \systemname will automatically extract the experimental procedures of the expected synthetic reactions from the paper. 
In addition, we provide a zoom-in button that, when clicked, opens a dialog containing an enlarged scatter plot. In this dialog, we provide two sliders: one to adjust the count of displayed points and another to modify the ``perplexity'' parameter in the t-SNE algorithm, which controls whether the algorithm emphasizes global patterns or local structures. We also include a tooltip to inform users about this parameter.

\subsubsection{Synthetic Reaction View}
The Synthetic Reaction View (Fig.\ref{fig:figure1}(C)) is designed to present the synthetic details extracted from the chosen paper. We allow the users to assess the experimental procedure in three aspects: the availability of necessary raw materials and reagents, the complexity of the experimental operations, and the difficulty in acquiring the necessary laboratory equipment (Fig.\ref{fig:figure1}(${\rm C_1}$)).

\textbf{Description.}
We use RDKit.js\footnote{https://www.rdkitjs.com/} to display the molecular structures of the reactants and products (Fig.\ref{fig:figure1}(${\rm C_2}$)) and present the experimental procedure of the expected reactions the users have specified extracted from the chosen paper in the form of a table, including \textit{reactants}, \textit{products}, \textit{solvents}, \textit{catalysts}, \textit{reagents}, \textit{reaction time}, \textit{instruments}, \textit{operation}, \textit{purification}, and \textit{yield}.
We provide links to these papers, allowing users to access them online with a single click.
Meanwhile, we provide a essential metric \textit{Context Relevancy}\cite{ragas2023Shahul} to aid users in assessing the trustworthiness of the extracted information (Fig.\ref{fig:figure1}(${\rm C_3}$)). We provide a view similar to a bullet chart to assist users in evaluating the effectiveness of information extraction. In this view, the narrower, darker bar represents the average Context Relevancy value of all extracted information with the wider, lighter bar representing the quartile range. The black vertical line indicates the value for this particular extraction result.
If users find the extracted results unsatisfactory, they have the flexibility to modify the results based on the article's content and expertise (Fig.\ref{tab:table1}).

\textbf{Interaction.}
By clicking the \textit{``Set Difficulty''} button, we allow the users to assess the experimental procedure in three aspects: \textit{material}, \textit{operation}, and \textit{equipment}. We categorize these into three discrete tiers of difficulty: challenging, moderate, and simple, with values 1 to 3 (Fig.\ref{fig:figure1}(${\rm C_1}$)). 
By recording the users' annotation of difficulty levels in three aspects and corresponding descriptive text, we utilize PubMedBERT to compute similarity scores, thereby providing recommendations for difficulties of experimental procedures that align with the users' predefined difficulty preference in subsequent annotations.
Then, the users can add the synthetic reaction to the end of the currently explored sequence after clicking the \textit{``Add into Route''} button. Additionally, we allow users to manually modify or add the extracted synthetic details after reviewing the article content or practical experimental results.

\subsection{Synthetic Route Construction View}
To help users construct and intuitively view multiple synthetic routes while making decisions based on various factors, we propose the Synthetic Route Overview in combination with the Rank View. Additionally, the Molecule Similarity View assists users in selecting the most promising intermediate molecules, guiding them away from less viable options during exploration. The Experimental Procedure Comparison View further support users in scrutinizing the synthetic information to make more informed choices.

\begin{figure}
    \centering
    \includegraphics[width=\linewidth]{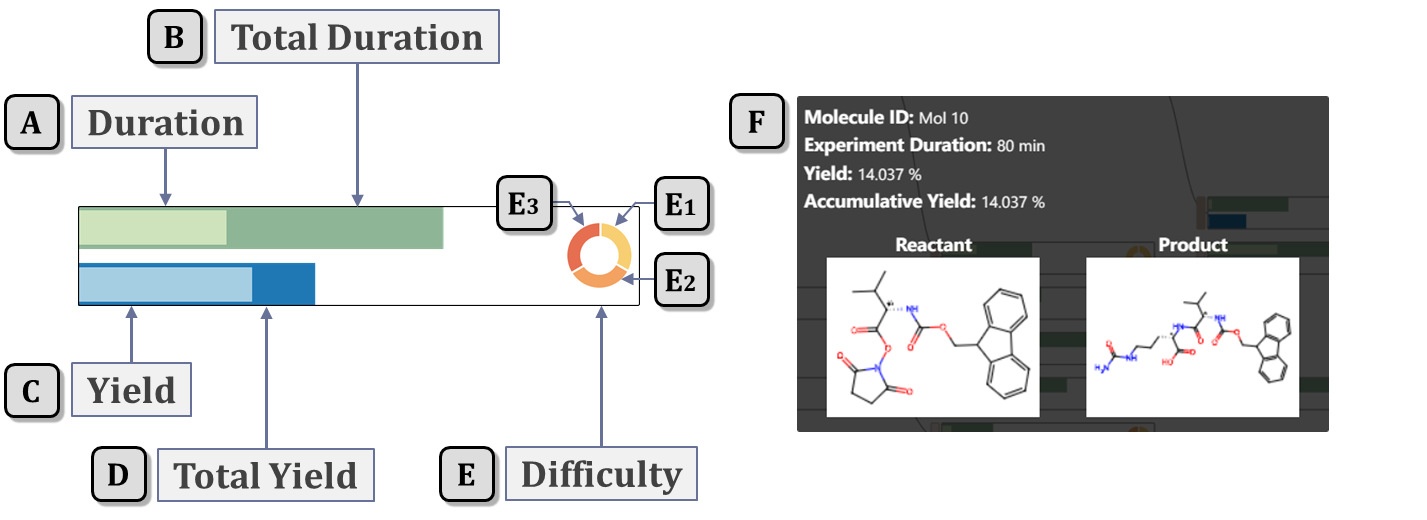}
    \caption{The design of the node glyph that represents a synthetic reaction: (A) and (B) respectively represent the duration and total duration accumulated from the root to this node. (C) and (D) respectively represent the yield and total yield. The donut glyph (E) presents the annotated experimental procedures' difficulty of this synthetic reaction. Specifically, (${\rm E_1}$), (${\rm E_2}$), and (${\rm E_3}$) correspond to the difficulty in acquiring reactants, obtaining and utilizing experimental instrument, and the complexity of the experimental operations, respectively. The tooltip (F) displays some details about this synthetic reaction.}
    \label{fig:figure4}
\end{figure}

\subsubsection{Synthetic Route Overview}
The Synthetic Route View (Fig.\ref{fig:figure1}(D)) proposes tree-form visualization to present decision sequences constituting synthetic routes, enabling users to explore various options of synthetic reactions.

\textbf{Description.}
In the synthetic route overview, the root represents the starting molecule of the whole synthetic route, and each subsequent node glyph (Fig.\ref{fig:figure4}) represents a synthetic reaction whose reactant is the product of the parent node. For instance, the product of \textit{Reaction 16} serves as the reactant of \textit{Reaction 27}.
The dual-toned bar on the upper side of each node glyph (Fig.\ref{fig:figure4}(A) (B)) represents reaction duration, with the lighter color representing the experimental duration of the current reaction and the darker color representing the cumulative total duration. The bar on the lower side (Fig.\ref{fig:figure4}(C) (D)) indicates yield, with the darker bar representing the yield of the reaction in the node and the lighter representing the cumulative total yield traced from the root to the current node.  
The donut glyph on the right side (Fig.\ref{fig:figure4}(E)) reflects the degree of difficulty in the three aspects: material acquisition (Fig.\ref{fig:figure4}(${\rm E_1}$), obtaining and utilizing experimental instruments (Fig.\ref{fig:figure4}(${\rm E_2}$), and the complexity of the experimental operations (Fig.\ref{fig:figure4}(${\rm E_3}$).
Nodes within the same decision sequence are connected by lines, while nodes at the same layer are vertically aligned.

\textbf{Interaction.}
When a node is hovered on, a pop-up tooltip (Fig.\ref{fig:figure4}(F)) shows the \textit{reaction ID}, which consists of the number of layers and the vertical ordering of the nodes, \textit{experiment duration}, \textit{yield}, \textit{total yield}, and the \textit{reactant} and \textit{product} of the corresponding synthetic reaction.

\begin{figure}
    \centering
    \includegraphics[width=\linewidth]{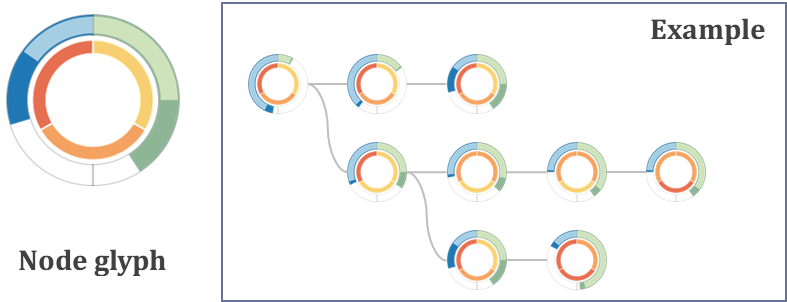}
    \caption{Design alternatives. The node glyph is designed to represent a synthetic reaction, which consists of two layers of arcs.}
    \label{fig:figure5}
\end{figure}

\textbf{Design alternatives.}
Fig.\ref{fig:figure5} illustrates alternative designs of the Synthetic Route Overview. 
The original design alternative consists of two layers of arcs, the outer ones representing the duration and yield, respectively, and the inner circle representing the difficulty of the experimental procedure.
However, the dual-layer arcs could be visually complex and potentially overwhelming, making it difficult for users to discern the key information quickly.
Moreover, comparing multiple nodes for yield or duration might not be intuitive, especially since these nodes are of the same sizes and there are many nodes in the view.
Therefore, we proposed the current design (Fig.\ref{fig:figure4}), which juxtaposes two bar charts while positioning a donut glyph on the right of the node, thereby facilitating user comparison across various nodes. Since the ``difficulty of a synthetic reaction'' are driven from text describing experimental procedures, it is challenging to quantify difficulties. So, we use a donut glyph, with three segments and color intensity representing three aspects of difficulties, which allows users to view the overall difficulty of the synthetic route while primarily making decisions based on duration and yield.

\textbf{Exploration.} Users usually adopt two kinds of exploration strategies using the Synthetic Route Overview, \ie, deep or bread exploration strategies. 
The deep exploration strategy involves constructing a single decision sequence from the starting molecule to the target molecule while the bread exploration strategy considers various possibilities to integrate multiple reactions.
Moreover, users often encounter multiple intermediates during the designing process, which may share similar structures, and must identify the most promising one for further exploration. To broaden the selection, users need to combine the deep and the broad exploration strategies to construct additional decision sequences that reach the step of producing similar intermediates. 

\subsubsection{Molecule Similarity View}
We design the Molecule Similarity View (Fig.\ref{fig:figure1}(F)) to help the users find potential intermediate molecules to perform subsequent explorations to obtain molecules that can be compared.
The first part of this view is the element left to each node glyph in the Synthetic Route Overview, which refers to the Molecule Similarity Mark (Fig.\ref{fig:figure1}(${\rm F_1}$)).
When the users click a node glyph, the color of all the Molecule Similarity Mark encodes the similarity between the product molecules of all other nodes in the tree and the chosen node's product molecule.
The other part of this view is the Molecule similarity Matrix (Fig.\ref{fig:figure1}(${\rm F_2}$)) right to the Synthetic Route Overview. 
Then, after the users choose a node in the Synthetic Route Overview and click the \textit{``Add into Comparison''} button, an additional column is added to the Molecule Similarity Matrix with a maximum number of five. 
The color intensity of each element in this column represents the molecular similarity between the product of this node and the products of all decision sequences.

\subsubsection{Rank View}
The users are required to rank various decision sequences represented by the routes from the root to the leaf in a tree structure considering the flexible weighting of multiple factors. Additionally, they need to observe each factor and the weighted sum to make informed decisions.

\textbf{Description.}
The horizontal coordinate of dot plots in the Rank View (Fig.\ref{fig:figure1}(${\rm E_2}$)) indicates the values of each factor, including \textit{the number of steps}, \textit{yield}, and \textit{duration}, which are the three most important factors in the decision-making of designing synthetic routes. 
The color and length of the bar on the right side of the view (Fig.\ref{fig:figure1}(${\rm E_3}$)) refer to the weighted total score for the three factors. 

Initially, we presented the scores for three attributes using bar charts. However, expert feedback indicated that presenting multiple bar charts together could lead to visual overload. Therefore, considering the need for ranking, visual simplicity, and quick task completion, we opt for dot plots. Moreover, the human eye is not as sensitive in visual perception to assessing length, \ie, the length of bars, as it is to dots.
We use bars with color intensity in the last column to represent the weighted total score. Because the bar chart uses multiple encodings, including the length of the bars, color intensity, and position on the common scale, it can represent the weighted total score while also allowing users to assess the rankings of each row intuitively.

\textbf{Interaction.} We provide a sorting feature that allows rows in the Rank View to be arranged based on their scores, facilitating users in selecting the optimal synthetic routes. When a user clicks on a row, the entire corresponding route in the Synthetic Route Overview will be highlighted, facilitating users' review.

\textbf{Legend.}
The legend of the Rank View presents the three factors, where the ratio of the length of each arc to the circumference of the corresponding circle represents the weight of each factor.
The users can edit the weights by clicking the button beside the legend (Fig.\ref{fig:figure1}(${\rm E_1}$)).

\subsubsection{Experimental Procedure Comparison View}
Before finalizing their decision on the specific synthetic route, the users can examine and assess the experimental procedures of several synthetic reactions(Fig.\ref{fig:figure1}(G)). By choosing a node within the tree and clicking the \textit{``Add into Comparison''} button, the users can compare the experimental procedure with other nodes in the form of a table. 
The examination facilitates the intuitive assessment of the difficulties in the experimental procedure of each synthetic reaction. 

\section{Case Study}

\begin{figure*}
    \centering
    \includegraphics[width=\linewidth]{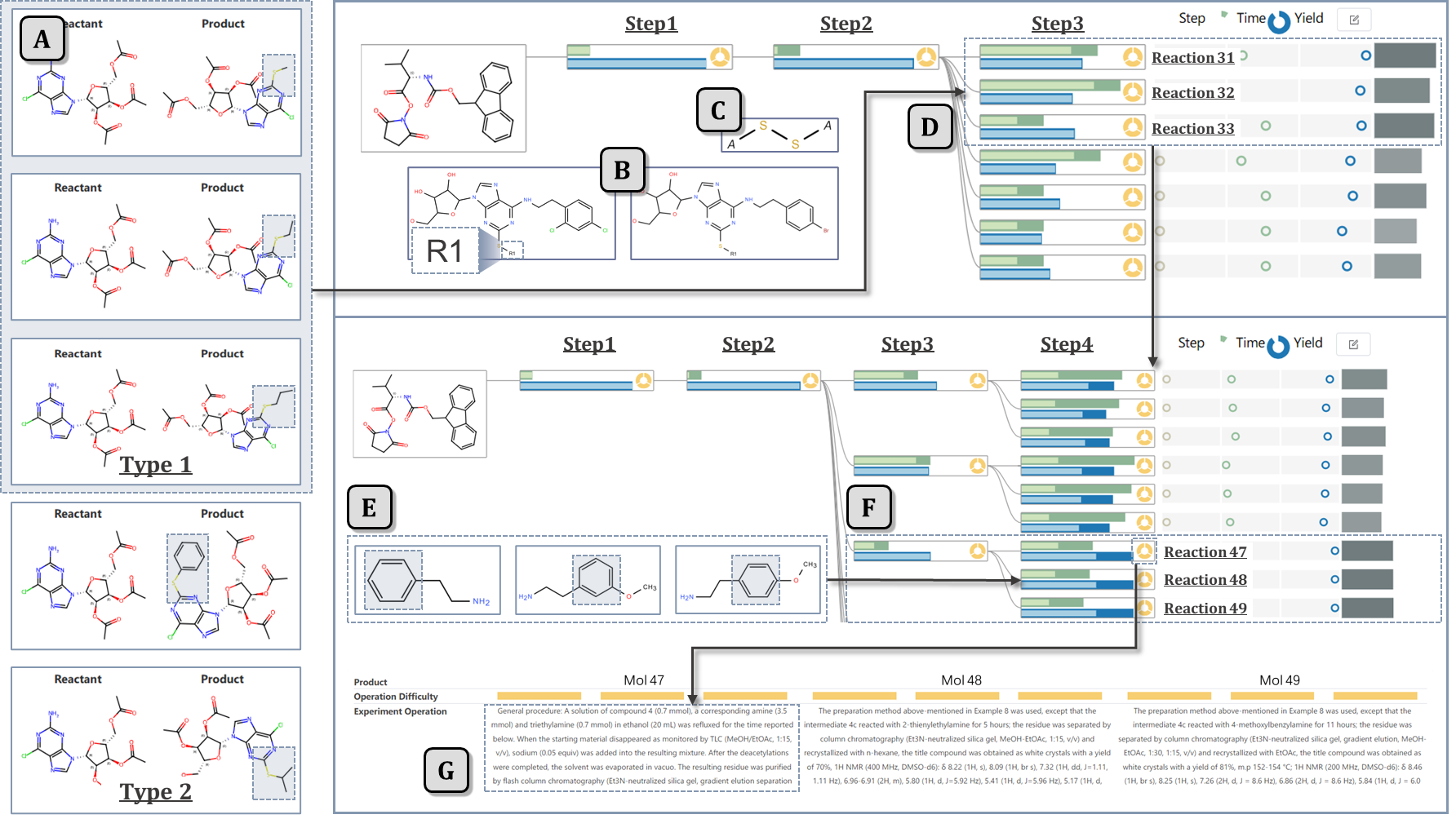}
    \caption{Design a synthetic route of a class of molecules in Case Study 2. 
    (A) presents several synthetic reactions integrated in \textit{Step3}. 
    (B) shows two of the target molecules. 
    (C) refers to a class of molecules containing double sulfur bonds (-S-S-). 
    (D) and (F) presents decision routes with relatively high weighted scores. 
    (E) presents molecules with similar structures as the reactants of the synthetic reactions in \textit{Step 4}.
    (G) details several experimental procedures of certain synthetic reactions, which have difficulty in experimental operations.
    }
    \label{fig:figure6}
\end{figure*}
We invited the interviewed experts to explore \systemname remotely online and recorded their exploration process.
Through these case studies, we summarized their insights to demonstrate the practical applications of our system.

\subsection{Case Study 1: Synthesis of Linker}
$\Ex{4}$, who specializes in anti-tumor targeting drugs, has been concentrating on Adcetris, a cancer treatment drug. He sought to use \systemname to design the synthetic route for Mc-Vak-Cit-PABC-PNP, a linker of Adcetris. He stated that the primary purpose of designing this synthetic route was to serve as a reference for experimental procedures intended for trials, not necessarily to actually produce the target molecules.

\textbf{User Input (R1).}
According to the existing research results and the actual laboratory situation, he chose the Fmoc-Val-OSu as the starting molecule (Fig.\ref{fig:figure1}(${\rm D_1}$)). 
He assigned initial weights to three factors: step count, duration, and yield. The weights for these factors were set as 0.1, 0.3, and 0.6, respectively, in (Fig.\ref{fig:figure1}(${\rm E_1}$)). \textit{``I am willing to accept long duration if it means achieving higher yields as often the significance in designing new synthetic route is the maximization of yield.''}
After observing the difference between the starting and target molecules, he found that the first step should be an amidation reaction based on his prior knowledge. Then, he inputted the name of the expected reaction in the Control Panel (Fig.\ref{fig:figure1}(${\rm A_1}$)).

\textbf{Combining deep and bread exploration strategies (R3).}
\systemname automatically retrieved papers related to the starting molecule and presented them in the Paper Projection View (Fig.\ref{fig:figure1}(B)). 
$\Ex{4}$ noted a point surrounded by a relatively long arc, indicating a high citation.
He checked the topic of this paper and found that this paper was about generating novel linkers for Adcetris.
Therefore, he determined this paper for further exploration. 
Once clicking the corresponding point in the view, \systemname automatically extracted several synthetic reaction information according to the inputted expected reaction whose reactant is the starting molecule, as (Fig.\ref{fig:figure1}(${\rm C_1}$)) shows. 
Adopting the bread exploration strategy, he integrated the identified candidate synthetic reactions into the Synthetic Route Overview (Fig.\ref{fig:figure1}(D)) following the root node. 
Subsequently, he wanted to explore if there were more candidates for the first step. Therefore, he began to explore two other papers.
Then, $\Ex{4}$ chose the most satisfied reaction \textit{Reaction 16} by checking the weighted score of these candidates in the Rank View to complete the decision sequence until the synthetic route reached the target molecule, \ie,  \textit{S1}.
$\Ex{4}$ indicated, \textit{``We need to construct a decision sequence from the starting molecule to the target molecule to ensure that this synthetic route is chemically feasible before subsequent full-scale exploration.''}
Once the synthetic route exists in principle, he can construct multiple decision sequences for comparison. 
It is worth noting that the product of different decision sequences may not be the same but may be structurally similar, and he can still compare these decision sequences.

\textbf{Choosing intermediate reactions for another exploration (R5).}
The first route \textit{S1} he completed had great difficulty in the experimental procedure, all of the experimental procedures in this route are difficult in experimental operations, so he decided to choose an intermediate synthetic reaction to explore other decision sequences. 
First, he chose an intermediate reaction \textit{Reaction 27} and constructed a decision sequence \textit{S2}.
Although \textit{S2} is not as difficult as \textit{S1} in experimental procedures, it has a significantly lower yield than \textit{S1}.
After checking the Molecule Similarity Mark in the (Fig.\ref{fig:figure1}(${\rm D_2}$)) and examining the Rank View, $\Ex{4}$ noted that \textit{Reaction 12} not only boasted a high weighted score but also had a product structure that is structurally similar to the product of \textit{Reaction 16}. 
This observation led him to construct a new decision sequence, ultimately choosing \textit{Reaction 12} for further exploration.
He started with \textit{Reaction 12} and constructed decision sequences such as those shown in (Fig.\ref{fig:figure1}(${\rm D_3}$)).

\textbf{Decision making (R4).}
After completing the construction of decision sequences, $\Ex{4}$ compared the structural similarity of \textit{Reaction 61}, \textit{62}, \textit{63}, \textit{55}, and \textit{56} in the Molecule Similarity View (Fig.\ref{fig:figure1}(${\rm F_2}$)) and found that the product of these reactions had the same main chain and only differed in side chains. 
$\Ex{4}$ indicated that when the product molecules of different sequences share structural similarities, they desire to compare these sequences and choose the optimal one. 
In such cases, he only needed to make minor adjustments to the chosen constructed route to adapt it for synthesizing the target molecule.
Consequently, he decided to rank the synthetic route by checking the Rank View (Fig.\ref{fig:figure1}(${\rm F_2}$)) and taking the difficulty of the experimental procedure into account. 
He thought that although \textit{S1} had a higher weighted score in the Rank View, the experimental procedure of this route was more difficult. So he chose \textit{S3}. 
\textit{``Although \textit{S3} did not have the highest total score, its experimental procedures are relatively moderate in difficulty. This indicates that we can try hands-on in the real experiment.''}

In this case, $\Ex{4}$ explored \systemname using flexible strategies, balancing various factors to guide decision-making, and constructing decision sequences in about 50 minutes, which results in the choice of a practical and promising synthetic route for a novel molecule.

\subsection{Case Study 2: Synthesis of A Class of Molecules} \label{sec:sec6.2}
Purine is a type of organic molecule that is found in nucleic acids and plays important roles in the human body, especially in fighting against viruses and bacteria.
$\Ex{2}$ had designed several new molecules based on purine, as (Fig.\ref{fig:figure6}(B)) shows, in which \textit{R1} refers to any chains only with carbon and hydrogen atoms. 
Since these molecules are newly designed, he needs to refer to the synthetic route of similar molecules to design the synthetic routes of these novel molecules. 
Based on the results of his earlier experiments, $\Ex{2}$ set guanosine as the starting molecule.
Considering the structural gap between the starting molecule and the target molecule, the expert specified that there would be four steps in this synthetic route.

\textbf{Compare intermediate synthetic routes (R2, R4).}
After completing the first two steps, $\Ex{2}$ started to examine papers to complete the third step of his construction process.
He found that the product of \textit{Reaction 21} could react with various molecules containing a disulfide bond (-S-S-)(Fig.\ref{fig:figure6}(C)) according to the extracted information of synthetic reactions from papers.
Based on his domain knowledge in chemistry, $\Ex{2}$ realized there are two possibilities for the main chain of a molecule containing a disulfide bond: either the main chain is straight (\textit{type1}) or the main chain branches out into branched chains (\textit{type2}). As a result, he explored six different papers in the Paper Projection View and the Synthetic Reaction Detail to integrate the synthetic reactions with seven different disulfide bond-containing reactants into the Synthetic Route Overview (Fig.\ref{fig:figure6}(D)). 
Each of these reactions can produce molecules that share a common main chain. 
He assessed these intermediate synthetic routes considering yield and duration in the Rank View and found that when the main chain of the molecule containing the disulfide bond used in the third step has a straight chain, the yield and reaction time are superior to those of the molecule containing a branched chain (Fig.\ref{fig:figure6}(A)). 
So, he chose \textit{Reaction 41}, \textit{42}, and \textit{43} to complete the decision sequences.
In the fourth step, he found that these reactants could react with a variety of molecules that shared similar structures (Fig.\ref{fig:figure6}(E)). 
As a result, the products of these reactions and the target molecules are similar in structure. After having constructed the decision sequences, he checked the Rank View and found the reactions in (Fig.\ref{fig:figure6}(F)) had the highest scores.

\textbf{Compare the Experimental Procedure (R2).}
Then, he chose the last three nodes in the fifth step of the tree in the Synthetic Route Overview to compare their experimental procedures. 
He found that the experimental method of \textit{Reaction 57} required the construction of a vacuum environment during evaporation (Fig.\ref{fig:figure6}(G)), which was relatively difficult to achieve.
Therefore, he chose to refer to the experimental procedures of \textit{Reaction 58} and \textit{59}.

The case demonstrates the importance of domain knowledge in assessing the choices during the designing process.
Additionally, the case indicates that the users need to compare textual data on several experimental procedures to ensure the chosen reactions are practically feasible, particularly avoiding complex operations. $\Ex{2}$ successfully completed this case within 35 minutes.

\section{Evaluation}
\subsection{Automatic Extraction Evaluation}

We conducted an experiment to validate the accuracy of the automatic extraction method of our system, introduced in Section \ref{sec:sec3}. 
We randomly collected 100 papers proposing synthetic routes, published from 2023 to the present, from four major chemical journals: \textit{The Journal of Organic Chemistry}, \textit{Organic Chemistry Frontiers}, \textit{Organic Process Research \& Development}, and \textit{European Journal of Organic Chemistry}. We invited four domain experts, each with over four years of experience in organic chemistry, to annotate the papers. We assigned 50 papers to each expert, ensuring each paper was annotated by two experts. Experts were asked to annotate the reactants, products, and yields of the synthetic routes. It is worth mentioning that experts may give multiple annotations to a paper if it proposes more than one synthetic route.
Then, we manually checked each annotation for conflicts. When finding conflicts, we consulted the opinions of the most experienced expert among those we invited to determine the correct annotation.
Next, we used the method in our system to extract the synthesis routes from each paper. 
Using the current standard evaluation metrics for NLP information extraction tasks\cite{ansari2023eunomia}, we employed Precision, Recall, and F1 Score to evaluate the accuracy of our method in extracting synthetic details from chemical papers. An extraction result was considered a \textit{True Positive} only if the reactants, products, and yields were all correctly identified. Any extraction result with at least one error was considered a \textit{False Positive}, and those missed in the extraction but present in the annotations were considered \textit{False Negatives}. The Precision was defined as the \textit{True Positives} out of all extracted details, indicating the accuracy of the method. The Recall was defined as the ratio of \textit{True Positives} to the total number of actual annotations, measuring the ability of the method to find information from papers. The \textit{F1 Score} was the harmonic mean of Precision and Recall, which is a comprehensive assessment of method performance.
Additionally, we tested these metrics using common information extraction tools for chemical papers: ChemDataExtractor\cite{cde22021mavracic}, ReactionDataExtractor\cite{reaction2022wilary}, and ChemRxnExtractor\cite{chemrxn2022guo}. 
Table \ref{tab:table2} shows the performance comparison among our system and common information extraction tools in chemistry. The precision, recall and F1 scores of our method are higher than those of the other methods.

\begin{table}[!htbp]
\centering
\caption{Comparison of our method and common information extraction tools in extracting information from chemistry papers}
\label{tab:table2}
\resizebox{0.46\textwidth}{!}{%
\begin{tabular}{lccc}
\hline
                      & \textbf{Precision} & \textbf{Recall} & \textbf{F1 Score} \\ \hline
ChemicalDataExtractor\cite{cde22021mavracic} & 0.871 & 0.527 & 0.653                  \\
ReactionDataExtractor\cite{reaction2022wilary} & 0.722 & 0.436 & 0.544                 \\
ChemRxnExtractor\cite{chemrxn2022guo}      & 0.942 & 0.627 & 0.753                 \\
Our System                 & \textbf{0.944} & \textbf{0.798} & \textbf{0.865}                 \\ \hline
\end{tabular}%
}
\end{table}
\subsection{User Study}
To validate the efficacy of \systemname, we conducted a semi-structured interview with the previous experts ($\Ex{1}$-$\Ex{6}$) mentioned in Section \ref{sec:sec3}. Besides, we also involved four new experts ($\Px{1}$-$\Px{4}$) in our evaluation, with more than four years of experience in organic chemistry. 
Each interview lasted one hour and a half.
We first introduced the background of our study (10 minutes). Then, we introduced our system in detail (15 minutes). 
Next, experts could familiarize themselves with the system through free exploration (15 minutes).
Then, they were asked to design a synthetic route using our system according to their individual requirements (30 minutes). 
Finally, we had a semi-structured interview with them to collect their feedback (20 minutes). 
After the interview, we asked the participants to complete a questionnaire multiple choice using a 5-point Likert scale to evaluate the effectiveness of \systemname in view representation, workflow enhancement, and overall usability. 
We documented their findings and comments on our system during the process.

\textbf{Workflow and System Performance.}
All the experts involved in the evaluation of \systemname expressed their appreciation for its workflow, as it aligned closely with their prior analysis process in organic chemistry. 
$\Px{1}$ mentioned, \textit{``The system can accelerate my design process significantly. Specifically, it makes it possible for me to compare different routes even if I have not completed them, resulting in substantial time savings.''} 
Specifically, we found that experts usually prioritize exploring potential synthetic reactions as much as possible in the early steps of their design process.
Then, they will identify the most promising one to conduct the subsequent exploration to complete a synthetic route. 
$\Ex{2}$ said that our ranking view provided him with effective insights to determine which one deserves further exploring.
Furthermore, $\Px{3}$ appreciated our system's capability to support retrospective exploration, recognizing that there might be several potential synthetic reaction candidates in the earlier exploration steps that could lead to improved synthetic routes.
\textit{``After completing a particular route, I often review previous steps to identify any new possibilities. The system is particularly helpful in quickly identifying these potential alternatives through molecular similarity analysis.''}
Particularly, $\Ex{5}$ praised the tree-form visualization for exploration, and he mentioned that this component is also suited for retrosynthetic tasks.
\textit{``In retrosynthesis, chemists employ a recursive approach, systematically working backwards from a target molecule to identify feasible synthetic routes. I can also construct a retrosynthetic route by inputting the target molecule through the tree-form visualization of \systemname.''} 

\textbf{Visualization and Interaction.}
$\Ex{1}$, $\Ex{2}$, and $\Px{2}$ all mentioned that the Paper Projection View provides an intuitive visualization of the distribution of retrieved papers, making it easy to identify which papers need to be examined first. 
Moreover, $\Ex{3}$ and $\Px{1}$ also highlighted the usefulness of allowing them to specify expected reactions and extract information from the papers quickly.
Additionally, $\Px{4}$ specifically pointed out that the Synthetic Route View combined with tooltips can effectively present the necessary information due to its straightforward design.
Finally, $\Px{3}$ and $\Px{4}$ pointed out the benefits of combining point visualizations with bar charts in the Rank View. They noted that this combination effectively highlighted the most crucial information (\ie, the weighted score displayed through the rightmost bar charts).
In summary, all the experts provided positive feedback on various aspects of the system's visualizations and interactions. 

\textbf{Suggestions.} 
The experts expressed their willingness to use the system but suggested some improvements. Specifically, they noted that the system currently only retrieves academic papers on chemical synthesis. Incorporating the ability to search for patents of this domain would broaden the coverage of synthesis information sources, thereby enhancing the usability of \systemname.
$\Px{1}$ indicated that experts in the field sometimes retrieve relevant papers by inputting multiple molecules with similar structures. 
Therefore, $\Px{1}$ suggested that \systemname can recommend structurally similar molecules and allow for the inclusion of more than one starting molecule. 
This approach would enhance the informativeness of the constructed synthetic routes.
Moreover, $\Px{4}$ suggested that we could take the price of raw material into account, which is also important in deciding whether to integrate a synthetic reaction.
\begin{figure}
    \centering
    \includegraphics[width=\linewidth]{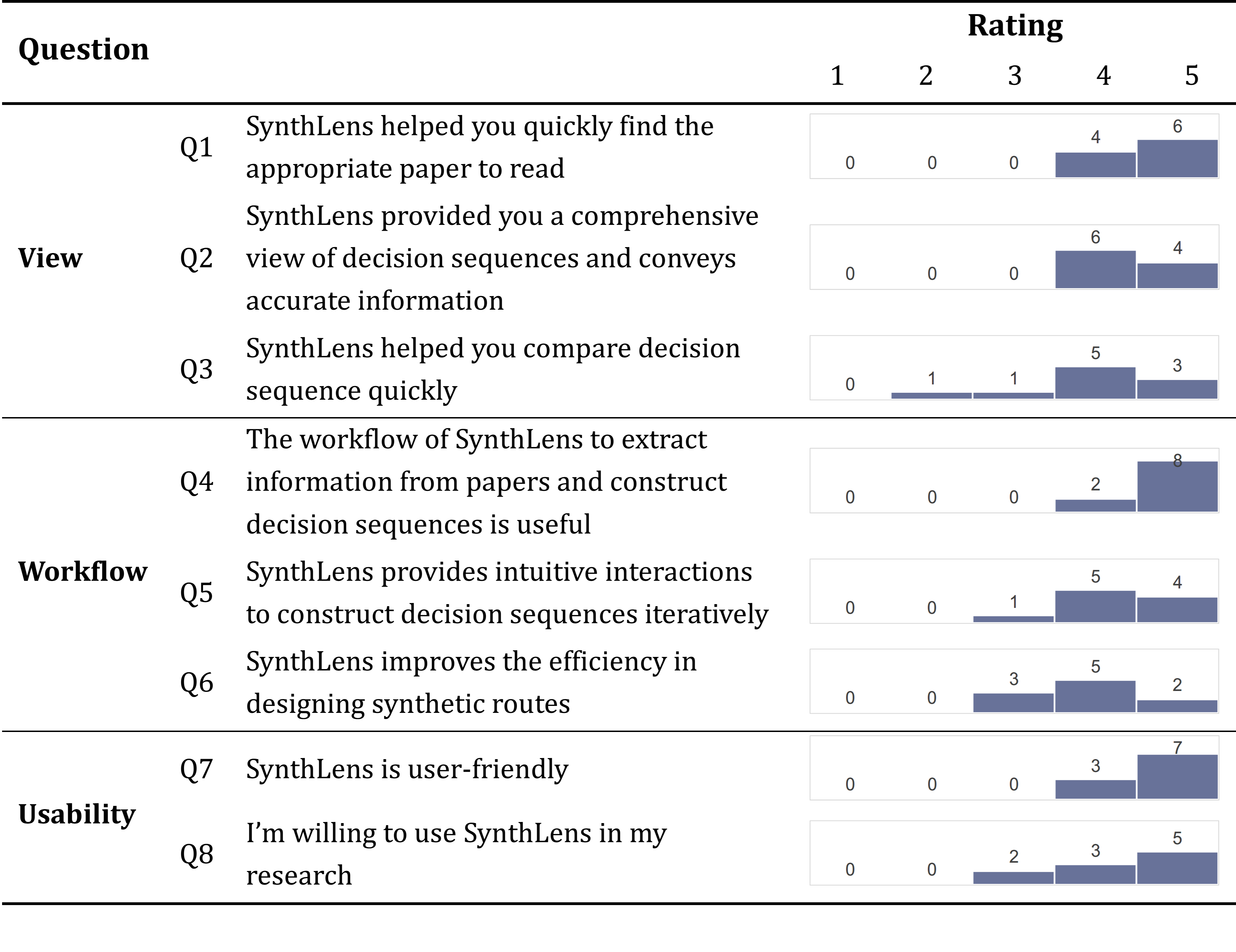}
    \vspace{-0.7cm}
    \caption{Questionnaire results. We designed our questionnaire in three aspects: the system's view, workflow, and usability. Experts are asked to rate by a 5-point Likert scale (from strongly disagree to strongly agree).}
    \label{fig:figure7}
\end{figure}
\section{Discussion}

\textbf{Runtime Performance.}
\systemname uses the PubMed API for paper retrieval, followed by the conversion of the paper content into high-dimensional embedding representations and dimensionality reduction. This is followed by the information extraction through LLM, which costs almost 1 minute or more and contributes to the main time consumption of the analytic process. However, \systemname' analysis process is designed in parallel mode, allowing users to continue exploring other papers while awaiting the results of the current analysis, thus ensuring the continuity and efficiency of the research process.
Notably, \systemname markedly reduces the time consumption and preserves the essential element of manual decision-making in comparison to traditional expert-driven approaches, where the expert manually searches, filters and extracts information from papers, subsequently constructing synthetic routes. 

\textbf{Design Implications.}
In our interviews, experts indicated that they can adopt the proposed workflow for designing synthetic routes in practice, treating the process as constructing a tree structure where a node represents a synthetic reaction, and the path from root to leaf defines the synthesis route, with attributes like cumulative yield and synthesis time to help identify the optimal route.
Moreover, we observed that experts frequently review intermediate nodes to explore promising synthetic route completions.
To facilitate this, \systemname includes the Molecule Similarity View, allowing users to select any node and automatically receive alternative possibilities of structure-similar molecules. 
Additionally, users' feedback indicates that our workflow is suitable for retrosynthetic tasks, where chemists deconstruct a target molecule to find feasible synthetic routes\cite{retrosynthesis2022ishida}. Our method may also apply to various multi-criteria decision-making scenarios, involving multiple options at each stage and enables users to efficiently evaluate the impact of choices on the outcome, such as fisheries management\cite{Maryam2012Fishery}, architecture construction\cite{JATOESPINO2014Construction}, and clinical decision-making\cite{Dolan2010Clinical}. For instance, in highway route selection, involving multiple objectives like safety, cost \etc, our tree-form visualization can help decision-makers compare and rank construction segment plans\cite{kalamaras2000highway}.

\textbf{Scalability.}
\systemname provides a tree-form visualization to assist experts in constructing and exploring synthetic routes. 
Although the constructed routes can usually be fully displayed within the current interface size, exploring the entire routes becomes challenging when the node number increases. To address this, our system supports zooming in and out to facilitate exploration, with corresponding adjustments in the size of the items in the Molecule Similarity View and the Rank View. Additionally, we provide a sorting feature that ranks various paths, making it easier for users to select the optimal synthetic route.
In the future, we plan to enhance the Synthetic Route Construction View with an overview feature that presents a summary of the tree-form visualization, which will facilitate navigation through complex options even when dealing with a large number of nodes.

\textbf{Limitations and Future Work.}
There are several areas that can be addressed to enhance \systemname further. 
Firstly, our system supports experts in specifying one currently starting molecule. However, expanding the capability to support multiple starting molecules would give experts more flexibility. 
Additionally, we can also recommend relevant papers about molecules similar to the current key molecule. This feature is valuable because molecules with similar structures, even if they differ slightly (such as in side chains), may share similar reaction conditions or types. By including these related molecules, the system can help users explore a wider range of relevant synthesis routes. Finally, incorporating additional information from papers, such as spectrograms and validation methods, would enrich the decision-making context for experts. 
Our main contribution is the proposed analytics analytic pipeline, which allows integration of advanced tools in place of our method, making it both adaptable and scalable to accommodate future advancements.

\section{Conclusion}
We introduce \systemname, a visual analytics system designed to assist organic chemists in designing synthetic routes by choosing optimal options at each step.
The novel tree-form visualization of \systemname simplifies understanding of decision sequences and helps optimal routes with the Rank View. 
The collaboration with domain experts helps us to establish design requirements and obtain valuable user feedback to refine \systemname.
Two case studies demonstrate the practical benefits and effectiveness of \systemname in practical applications. We also envision expanding the application of \systemname to other decision-making scenarios in chemistry fields and scientific researches.

\section*{Acknowledgments}
We would like to thank the reviewers and all participants in our studies for their valuable input. The research was partially supported by the National Natural Science Foundation of China (Grant No. 62172289).


 

\begin{IEEEbiography}[{\includegraphics[width=1in,height=1.25in,clip,keepaspectratio]{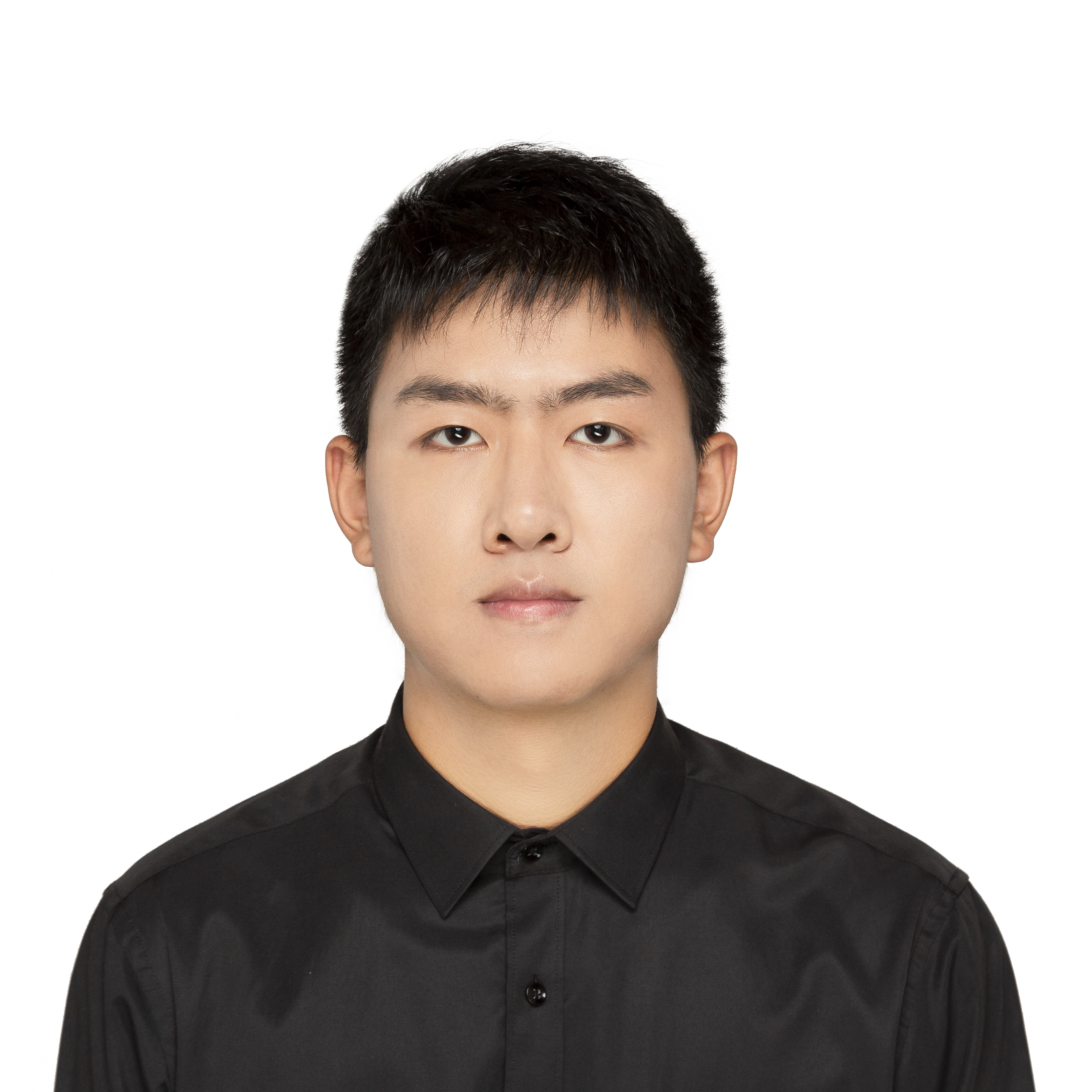}}]{Qipeng Wang} received his BS in computer science from Sichuan University in 2023. He is currently pursuing a master's degree in the College of Computer Science at Sichuan University. His research interests include visualization and visual analytics.
\end{IEEEbiography}

\vspace{-33pt}

\begin{IEEEbiography}[{\includegraphics[width=1in,height=1.25in,clip,keepaspectratio]{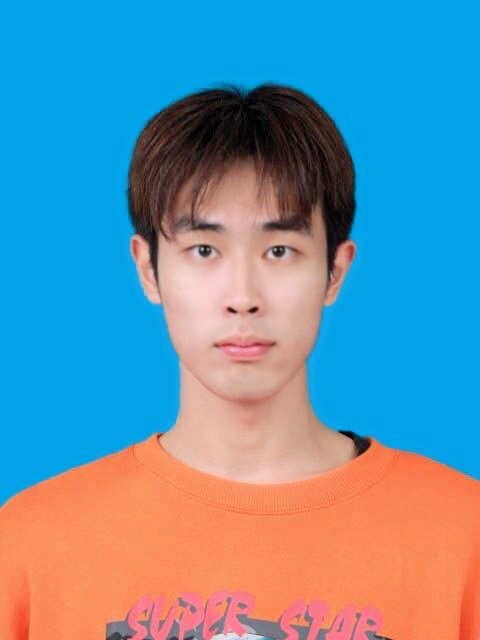}}]{Rui Sheng} is currently a third-year Ph.D. candidate in the HKUST VisLab, within the Department of
Computer Science and Engineering at the Hong Kong University of Science and Technology. Under
the guidance of Professor Huamin Qu, his research focuses on data visualization, human-AI collaboration, and decision-making, particularly in critical domains such as healthcare. For more information, please visit \href{https://dylansheng.github.io/}{https://dylansheng.github.io/}.
\end{IEEEbiography}

\vspace{-33pt}

\begin{IEEEbiography}[{\includegraphics[width=1in,height=1.25in,clip,keepaspectratio]{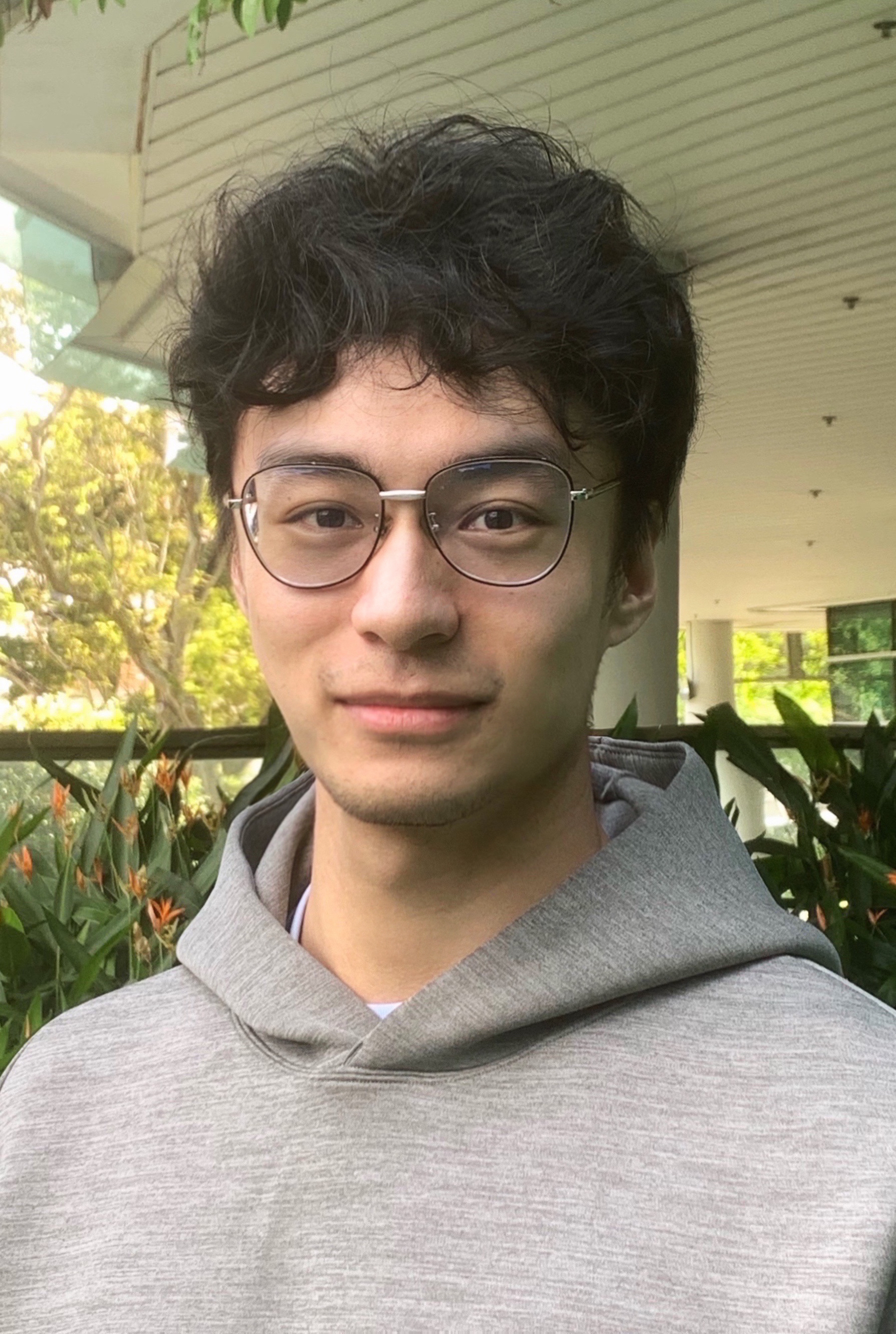}}]{Shaolun Ruan} is currently a Ph.D. candidate in School of Computing and Information Systems at Singapore Management University (SMU). His work focuses on developing novel graphical representations that enable a more effective and smoother analysis for humans using machines, leveraging the methods from Data Visualization and Human-computer Interaction. He received his bachelor’s degree from the University of Electronic Science and Technology of China (UESTC) in 2019. For more information, kindly visit \href{https://shaolun-ruan.com/}{https://shaolun-ruan.com/}.

\end{IEEEbiography}

\vspace{-33pt}

\begin{IEEEbiography}[{\includegraphics[width=1in,height=1.25in,clip,keepaspectratio]{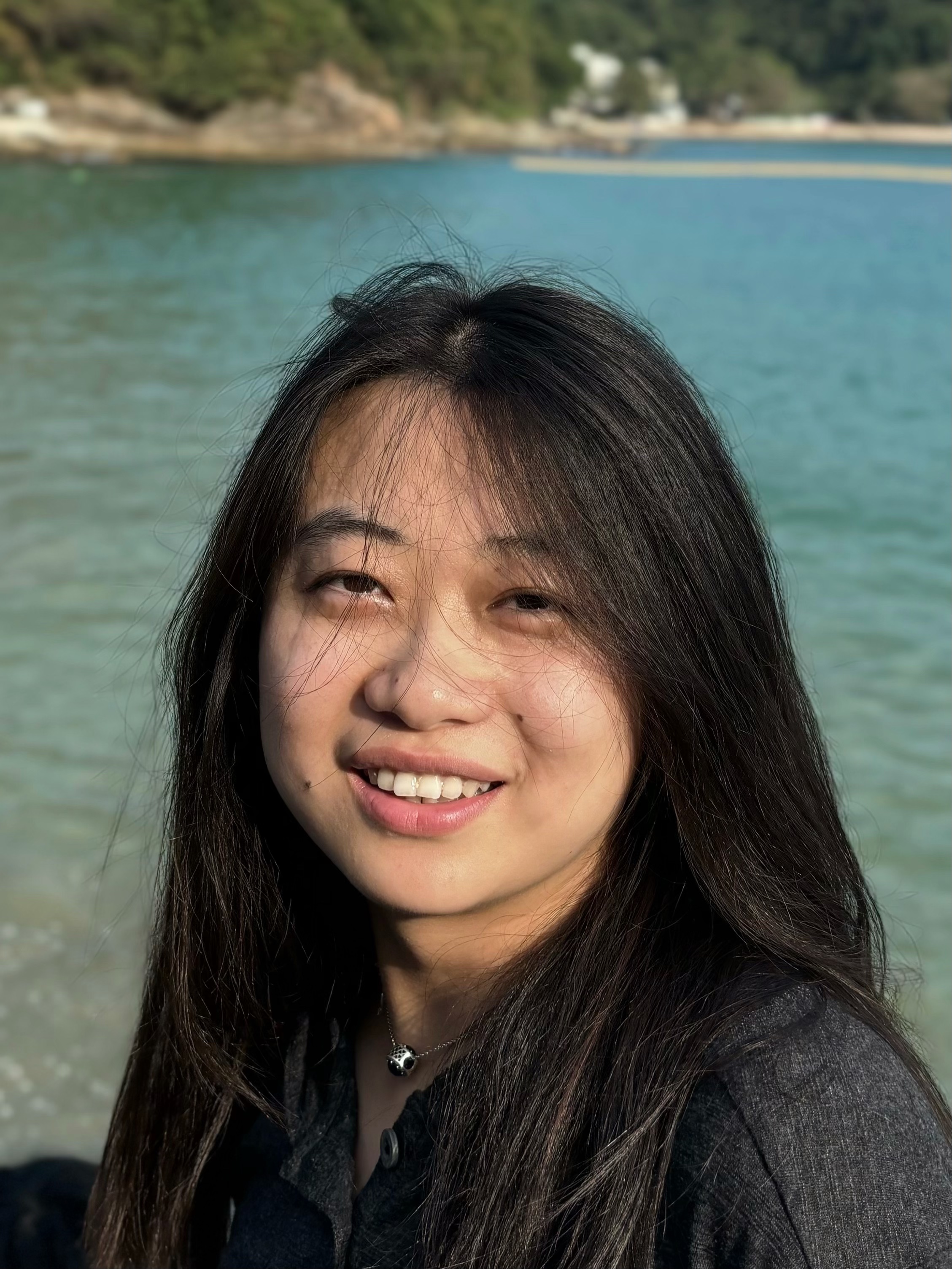}}]{Xiaofu Jin} is currently a third-year Ph.D. candidate in the HKUST VisLab, within the Department of Academy of Interdisciplinary Study at the Hong Kong University of Science and Technology. Her research focuses on accessibility, cognitive modelling, VR/AR, and health applications. For more information, please visit \href{https://xiaofu-jin.github.io//}{https://xiaofu-jin.github.io/}.

\end{IEEEbiography}

\vspace{-33pt}

\begin{IEEEbiography}[{\includegraphics[width=1in,height=1.25in,clip,keepaspectratio]{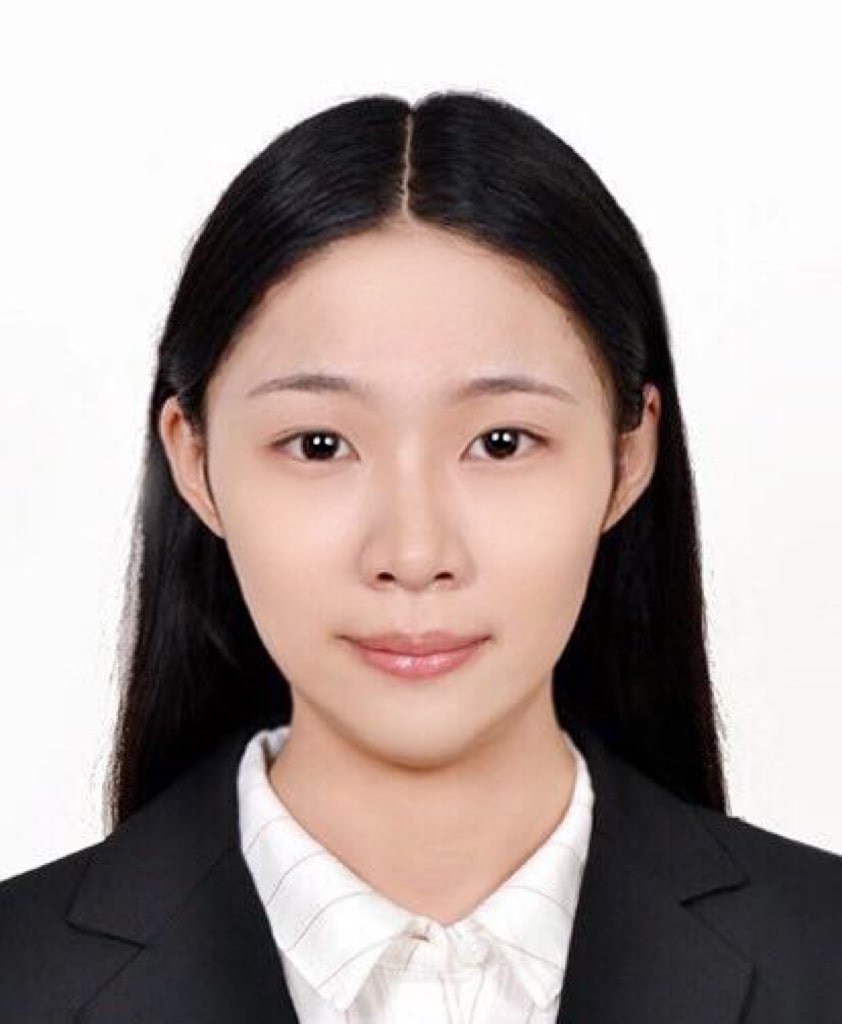}}]{Chuhan Shi} is an associate professor in the School of Computer Science and Engineering at Southeast University, China. She obtained her Ph.D. degree in Computer Science and Engineering from the Hong Kong University of Science and Technology in 2023. Her research interests include data visualization, visual analytics and human-computer interaction. For more details, please refer to \href{https://shichuhan.github.io/}{https://shichuhan.github.io/}.

\end{IEEEbiography}

\vspace{-33pt}

\begin{IEEEbiography}[{\includegraphics[width=1in,height=1.25in,clip,keepaspectratio]{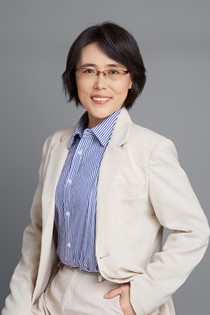}}]{Min Zhu}
received the PhD degree in applied mathematics from Sichuan University, in 2004. She is currently a professor at the College of Computer Science, Sichuan University, and has presided over a number of national and provincial research projects. Based on the research works, she has published more than 100 academic papers in journals and conferences. Her current research interests include visualization, visual analysis, and bioinformatics.
\end{IEEEbiography}

\vspace{11pt}


\vfill

\end{document}